\documentclass{aastex63}

\received{\today}
\revised{}
\accepted{}

\submitjournal{ApJL}

\shorttitle{Three-Body Resonances in the Saturnian System}
\shortauthors{{\'C}uk and El Moutamid}

\begin{document}

\title{Three-Body Resonances in the Saturnian System}

\correspondingauthor{Matija {\'C}uk}
\email{mcuk@seti.org}

\author{Matija {\'C}uk}
\affil{SETI Institute \\
339 North Bernardo Ave, Suite 200 \\
Mountain View, CA 94043, USA}

\author{Maryame El Moutamid}
\affil{Cornell Center of Astrophysics and Planetary Sciences, \\
Department of Astronomy and Carl Sagan Institute\\
Cornell University \\
326 Space Science Building\\
Ithaca, NY 14853, USA}

\begin{abstract}
Saturn has a dynamically rich satellite system, which includes at least three orbital resonances between three pairs of moons: Mimas-Tethys 4:2, Enceladus-Dione 2:1, and Titan-Hyperion 4:3 mean-motion resonances. Studies of the orbital history of Saturn's moons usually assume that their past dynamics was also dominated solely by two-body resonances. Using direct numerical integrations, we find that three-body resonances among Saturnian satellites were quite common in the past, and could result in a relatively long-term, but finite capture time (10 Myr or longer). We find that these three-body resonances are invariably of the eccentricity type, and do not appear to affect the moons' inclinations. While some three-body resonances are located close to two-body resonances (but involve the orbital precession of the third body), others are isolated, with no two-body arguments being near resonance. We conclude that future studies of the system's past must take full account of three-body resonances, which have been overlooked in the past work.
\end{abstract}


\section{Introduction} \label{sec:intro}

Three-body orbital resonances must include three different bodies orbiting the same central object. The potential associated with this resonance experienced by any of the three bodies will be proportional to masses of the each of the other two participants, necessarily making three-body resonances (TBRs) weaker than the two-body kind, at least in systems dominated by the central body. The 1:2:4 orbital period relationship between Jupiter's moons Io, Europa and Ganymede was the first known three-body resonance \citep{md99}. This ``Laplace resonance'' involves three bodies being in two simultaneous two-body resonances; chains of two body resonances are also seen in some exoplanet systems \citep{fab10, goz16, mil16, lug17, mor20, sie21}. Some exoplanet systems are found in TBRs of the zeroth order \citep{mcd16, gol21}, meaning they have an argument involving only mean longitudes, without any secular angles. Zeroeth order three body resonances should, in principle, not affect eccentricities and inclinations, and their dynamics is different from eccentricity-type TBRs discussed in this Letter.

Three-body resonances have been studied in some detail in the main asteroid belt \citep{nes98, mor99}, where they are a major source of dynamical chaos. Similarly, in closely-packed planetary or satellite systems, this type of resonances have been found to be an important source of chaos \citep{qui11, qui14, pet20}. Three-body resonances are in many systems overlapping with each other, or have overlapping subresonances, making chaos inevitable. Among planetary satellites, \citet{cuk20} found that the moons of Uranus may have experienced capture into three-body resonances in the past. Currently, Uranian moons Miranda, Ariel and Umbriel are close to a zeroth-order three body resonance \citep{gre75}, but this resonance is not exact and may become exact in the future (or may have been exact in the past).

In the past work on the Saturnian system, researchers considered only two-body mean-motion resonances as candidates for resonance capture. Since two-body resonances include interactions between a pair of moons, it was often convenient to ignore the presence of other bodies in the system. In our own past work we considered possible past resonant chains of three moons, with Mimas, Enceladus and Dione having a 2:3:6 period ratios \citep{elm20}, a form of Laplace resonance, but more recent results suggest that this resonance is not consistent with the low inclination of Enceladus \citep{elm21}.  However, our recent numerical work has uncovered, rather unexpectedly, that the mid-sized icy moons of Saturn can be captured into {\it three-body} resonances when evolving at ``realistic'' tidal migration rates (i.e. with multi-Gyr orbital evolution timescales). We find three kinds of three-body resonances: semi-secular TBRs (Section \ref{sec:semisec}) in which two bodies are in a mean-motion resonance and interact secularly with a third, combinations of two-body resonances that do not constitute complete resonant chains (Section \ref{sec:chains}), and isolated TBRs (Section \ref{sec:isolated}), which are not in proximity to any two-body resonances.

\section{Semi-Secular Three-Body Resonances} \label{sec:semisec}

The first kind of the three body resonance we found in the Saturnian system are within second-order two-body mean motion resonances (MMRs). A good example is the Mimas-Tethys 3:1 MMR, which may have happened relatively recently in some reconstructions of the system's history \citep{md99}. When integrating this resonance numerically \citep[using the integrator {\sc simpl};][]{cuk16} in the presence of other moons of Saturn\footnote{All our simulations included Saturn’s oblateness moment $J_2$, Mimas, Enceladus, Tethys, Dione, Rhea and Titan, as well as the Sun (as a perturber of satellites) and Jupiter (pertubing Saturn's heliocentric orbit).}, we found that it contained more than six important sub-resonances. Conventionally, the 3:1 MMR between Mimas and Dione should have the three inclination sub-resonances (usually designated $i^2_M$, $i_M i_D$ and $i^2_D$, with subscripts $M$ and $D$ referring to Mimas and Dione, respectively) and three eccentricity sub-resonances \citep[$e^2_D$, $e_M e_D$ and $e^2_M$;][]{md99}. However, we found additional sub-resonances involving not just Mimas and Dione, but also other moons. The strongest of these are those associated with Titan: $i_M i_T$ sub-resonance with the resonant argument $3 \lambda_D - \lambda_M - \Omega_M - \Omega_T$, and $e_M e_T$ sub-resonance with the argument $3 \lambda_D - \lambda_M - \varpi_M - \varpi_T$ ($\lambda$, $\Omega$ and $\varpi$ are respectively the mean longitude, and the longitudes of the ascending node and the pericenter; subscript $T$ refers to Titan). Due to significant inclination and eccentricity of Titan, as well its large mass, we find these resonances to be significant, and that capture into $e_M e_T$ sub-resonance is assured if Mimas' eccentricity is low enough (we used an initial $e_M \simeq 10^{-4}$, based on the assumption that moons formed with very low eccentricities and inclinations). 

\begin{figure*}
\epsscale{.8}
\plotone{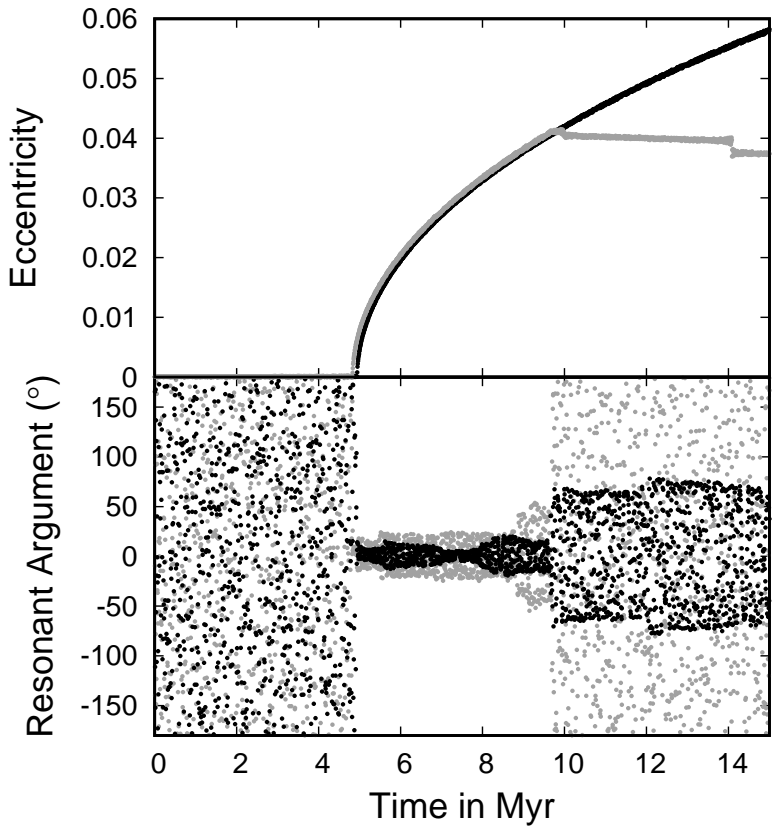}
\caption{Two simulations of capture into semi-secular three-body resonance between Mimas, Dione and Titan, illustrated by the evolution of the eccentricity of Mimas (top panel) and the resonant argument $ 3 \lambda_D - \lambda_M - \varpi_M - \varpi_T$ (bottom panel). In the simulation plotted in black both Mimas and Dione are assumed to have very low eccentricities ($e \leq 10^{-3}$) before the resonance was encountered, while Titan was assumed to be on its present orbit, but with eccentricity of e=0.033 (to account for subsequent eccentricity damping). The resulting three-body resonance persists until the end of the simulation.  In the simulation plotted in gray, Dione was assumed to initially have $e_D=0.01$, enhancing secondary resonances, seen in the resonant argument plot as jumps in the libration argument, and breaking the three-body resonance just before 10 Myr. The small kick to the eccentricity of Mimas at 10~Myr is due to the $3 \lambda_D - \lambda_M - \varpi_M - \varpi_D$ sub-resonance, while the larger kick at 14 Myr is due to one with the argument $3 \lambda_D - \lambda_M - 2 \varpi_M$.\label{semisec}}
\end{figure*}

Fig. \ref{semisec} plots two numerical simulation using {\sc simpl} of a part of the Mimas-Dione 3:1 MMR crossing, covering the encounters with the eccentricity-type sub-resonances. The crossing of the $e_D^2$ sub-resonance produces little effect, but before the system can reach the $e_M e_D$ sub-resonance, Mimas is captured into a $e_M e_T$ sub-resonance. This capture is robust and happens for all simulations that assume equilibrium tides $Q/k_2=4000$ for Saturn, and initial $e_M \simeq 10^{-4}$ (we ran several dozen). If we assume an initially $e \approx 10^{-3}$ orbit for Dione (simulation plotted in black), this resonance capture appears to continue indefinitely, and is likely to result in tidally-induced melting of Mimas which is inconsistent with its present appearance \citep{der88, nev17}. When we start the simulation with eccentric Dione ($e_D=0.01$), secondary resonances disrupt the $e_M e_T$ sub-resonance, leaving Mimas with $e_M \approx 0.04$, which allows Mimas to cross $e_M e_D$ and $e_M^2$ sub-resonances without capture. This sub-resonance is one possible origin mechanism for the observed eccentricity of Mimas ($e_M \approx 0.02$), assuming moderate eccentricity dissipation by tides within Mimas over the last 50-100 Myr \citep[cf.][]{mey08}.

We term the resonances of the type seen in Fig. \ref{semisec} ``semi-secular'' three-body resonances, as the third body (Titan) interacts solely through its secular elements with the two bodies that in a MMR (Mimas and Dione). This is in analogy with two-body semi-secular resonances\footnote{First use of this term in the Astrophysics Data System is by \citet{cel17}.} like the evection resonance \citep{tou98}, where the mean motion of one body is commensurable with the orbital precession of another. Semi-secular three-body resonances among giant-planet moons were first found by \citet{zha07} in their simulations of the dynamical past of the Neptunian system, where Triton's orbital plane was affecting inclination-type resonances between the inner moons. Unlike that case where an exterior perturber's plane affected the inclination-type two-body resonance, in this case we have Titan affecting both inclination-type and eccentricity-type sub-resonances of the Mimas-Dione 3:1 MMR, as Titan is both eccentric and inclined. The eccentricity sub-resonance $e_M e_T$ typically results in a capture, while inclination sub-resonance $i_M i_T$ does not for the tidal parameters we used, possibly because of relatively high-eccentricity of Titan's orbit. The semi-secular three-body inclination sub-resonance does give a kick to the inclination of Mimas, implications of which for the capture into the Mimas-Tethys 4:2 MMR will be addressed in future work.

In general, it is clear that all second-order two body resonances in the Saturnian System harbor additional three-body sub-resonances that have previously not been appreciated. Large oblateness of Saturn generally makes all the sub-resonances well-separated, suppressing chaos and, when the resonances are strong enough, enabling stable long-term capture \citep[except for small moons on very close orbits, see][]{elm14}. While the three-body semi-secular resonances with Titan as the secular perturber tend to be strongest, \citet{cuk16} have seen (but not realized the wider importance of) temporary capture of Tethys into a semi-secular resonance with Dione and Rhea with an argument $5 \lambda_R - 3 \lambda_D - \varpi_D-\varpi_{\Theta}$ (subscript $\Theta$ refers to Tethys). As the third (secular) participant in this kind of resonance can be a larger moon perturbing two smaller ones (as in Fig. \ref{semisec}), or a smaller one being perturbed by two larger ones \citep[Tethys in][Fig. 8]{cuk16}, the range of possible outcomes is vast, and only direct numerical simulations can determine which sub-resonances are dynamically relevant.

\section{Combinations of Two-Body Resonances}\label{sec:chains}

The term ``resonant chain'' is typically used to describe a group of orbiting bodies each of which is in a two-body resonance with the next (like Io, Europa and Ganymede). However, in our simulations of the dynamical evolution of the Saturnian system we found that two-body commensurabilities can produce a three-body resonance without making a classic resonant chain. In a classic resonance chain, there are multiple two-body resonances each of which has resulted in a capture; in the case we are discussing here one of the resonances is encountered divergently and could not by itself result in capture.

\begin{figure*}
\epsscale{.8}
\plotone{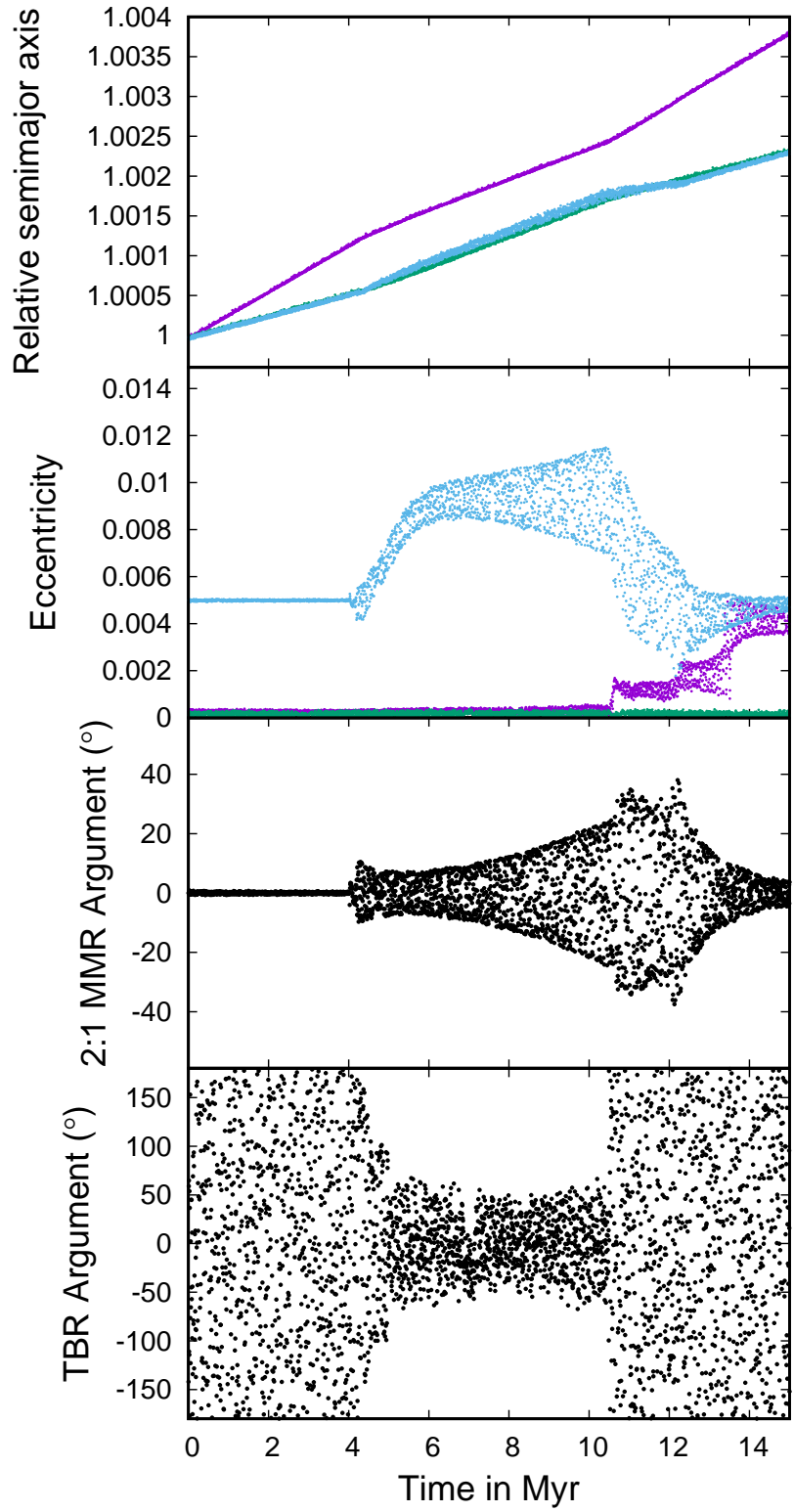}
\caption{A simulations of capture into a three-body resonance between Enceladus (light blue), Tethys (magenta) and Dione (dark green). The three-body resonant argument (bottom panel) is $4 \lambda_E - 11 \lambda_{\Theta} + 8 \lambda_D - \varpi_E$. This three-body resonance is a consequence of Enceladus and Tethys crossing their 11:8 MMR while Enceladus and Dione are locked in their 2:1 MMR (second panel from the bottom). Three-body resonance breaks at about 10.5 Myr, and the subsequent kicks to the eccentricity of Tethys are caused by various sub-resonances of the Eneceladus-Tethys 11:8 MMR. \label{chain}}
\end{figure*}

Fig. \ref{chain} shows a three-body resonance between Tethys, Enceladus and Dione, with the latter two also participating in their (current) 2:1 resonance. The resonant argument of the three-body resonance is $4 \lambda_E - 11 \lambda_{\Theta} + 8 \lambda_D - \varpi_E$ (where subscripts $E$ and $\Theta$ refer to Enceladus and Tethys, respectively), and the bottom panel of Fig. \ref{chain} shows a consistent libration of this angle after resonant capture at 4-5~Myr. This resonant argument is a combination of a pair of two-body commensurabilities: Enceladus-Dione 2:1 and Enceladus-Tethys 11:8 (with an argument $11 \lambda_{\Theta} - 8 \lambda_E - 3 \varpi_E$). While the former is a true stable resonance, the latter is a third-order MMR that is being crossed divergently, i.e. normally its crossing would not result in the resonance capture. However, by combining these two commensurabilities into a three-body resonance, a first order argument amenable to capture can be constructed. This is not really a resonant chain, as Tethys is not convergently evolving into a two-body resonance with another moon, but this TBR is still a combination of a pair of resonant two-body arguments.

While the TBR acts directly to increase the eccentricity of Enceladus, its eccentricity is also increased by Enceladus being pushed by the TBR deeper into the 2:1 MMR with Dione. The eccentricity growth is limited by strong eccentricity tides we assumed for tides within Enceladus \citep[$Q/k_2 = 100$;][]{lai12}. However, the eccentricity of Enceladus does not reach a new equilibrium, as two modes of resonant excitation appear to produce a growing oscillation in eccentricity, correlated with an increasing libration width of the three-body resonant argument. The three-body resonance eventually breaks and the libration amplitude of the Enceladus-Dione 2:1 MMR resonant argument gradually damps back to low values.

It is tempting to consider the possibility that this three-body resonance did happen in the past, possibly about 10 Myr ago, contributing to the past tidal heating of Enceladus. However, the probability of the resonant capture, assuming the present eccentricity of Enceladus, is only about 10-20\%, based on a limited number of our numerical simulations. Furthermore, the passage of Tethys through the many sub-resonances of its 8:11 MR with Enceladus, concurrent with the already established Enceladus-Dione 2:1 MMR, may be inconsistent with the low inclination of Enceladus and the survival of the Mimas-Tethys 4:2 resonance (which should have been established at this time). These and other constraints on the past evolution from Enceladus's inclination will be addressed in a separate paper \citep{elm21}. Here we use this example to demonstrate that Enceladus can be pushed deeper into the resonance with Dione by a three-body resonance, even when the third body (Tethys) is migrating divergently from Enceladus.

\section{Isolated Three-Body Resonances}\label{sec:isolated}

\begin{figure*}
\epsscale{.8}
\plotone{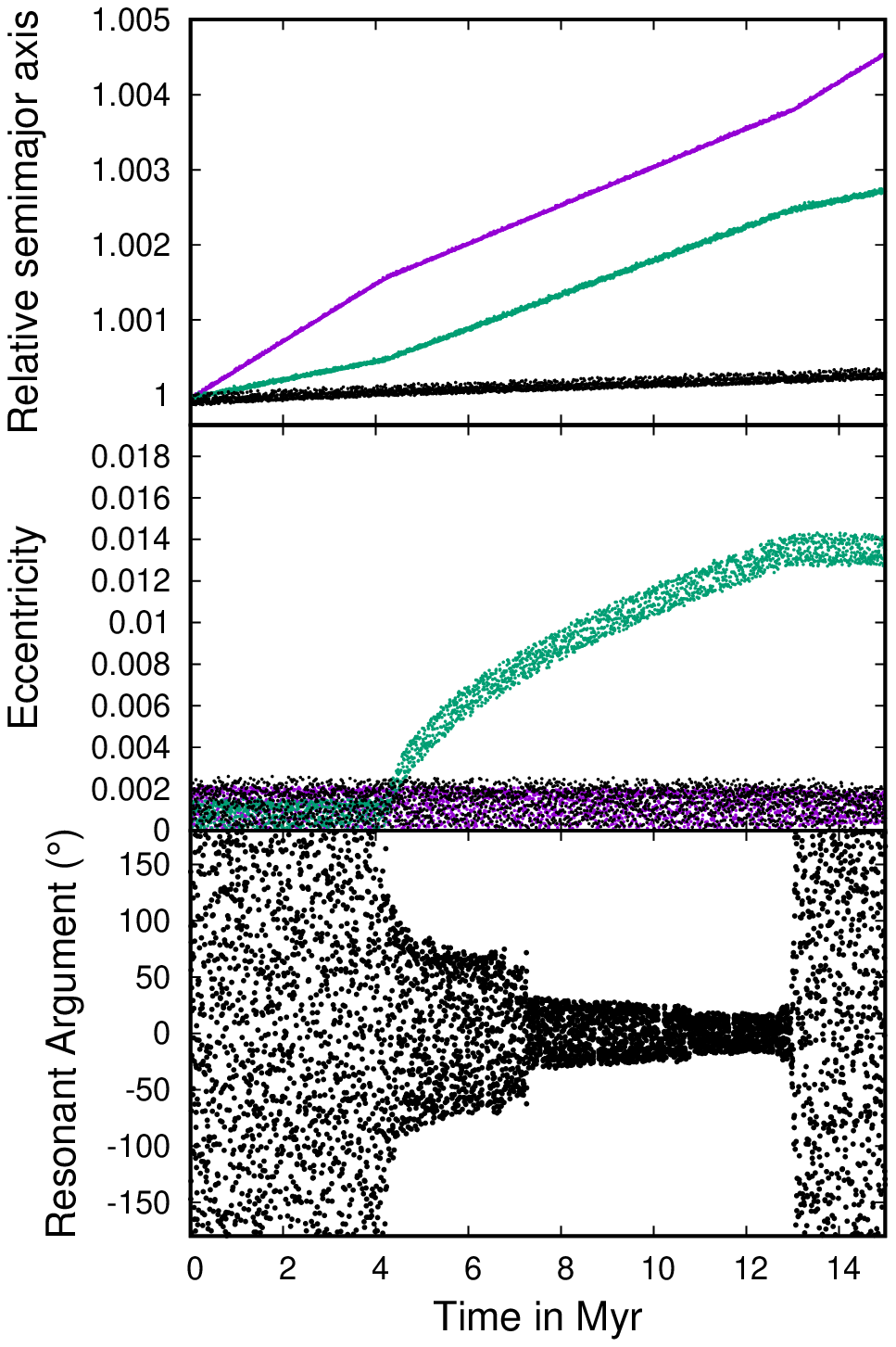}
\caption{A simulations of capture into an isolated three-body resonance between Tethys (magenta), Dione (dark green) and Rhea (black). The resonant argument, shown in the bottom panel, is $5 \lambda_D - 3 \lambda_{\Theta} - \lambda_R - \varpi_D$. If we assume that Saturn's moons are evolving though equilibrium tides, this resonance should have happened about 20 Myr ago. Note that Dione was given a low eccentricity ($e_D=0.001$) at the beginning of the simulation; present day eccentricity $e_D=0.0022$ usually leads to passage without capture. Depending on the eccentricity damping within Dione, this TBR could be the source of Dione's present eccentricity, or may have been passed without capture if Dione's eccentricity predates this resonance.\label{tbr1}}
\end{figure*}

The third kind of three body resonances we identify are not close to any two-body resonances, but are completely isolated in frequency-space. They typically involve slow-changing resonant argument, typically of the form $n_1 \lambda_1 + n_2 \lambda_2 + n_3 \lambda - \varpi_x$ where $n_1+n_2+n_3-1=0$ are integers and $x=$1,2 or 3. While the resonance shown in Fig. \ref{chain} has a three-body resonant argument of the similar type, in case of isolated three-body resonances there should be no resonant two body arguments (like $2 \lambda_D -\lambda_E - \varpi_E$ in Fig. \ref{chain}). Fig. \ref{tbr1} shows an instance of capture into an isolated three-body resonance, encountered serendipitously in our simulations of the system's long-term evolution. Tethys-Dione-Rhea resonance shown in Fig. \ref{tbr1} has an argument $5 \lambda_D - 3 \lambda_{\Theta} - \lambda_R - \varpi_D$ (where subscript $R$ refers to Rhea). This resonance changes the rates of orbital evolution of Tethys, Dione and Rhea, and can change the estimates of the timing of past dynamical events (and even system's age) based on tidal evolution rates of individual moons \citep{cuk16}. 

The closest match in the literature to these isolated three-body resonances are probably the past three-body's MMRs among the moons of Uranus found by \citet{cuk20}. As we noted in Section \ref{sec:semisec}, larger oblateness of Saturn appears to make three-body resonances among Saturn's moons better separated and therefore less chaotic than they are at Uranus. In general, we find that three-body resonances eventually break when another resonance is encountered by one of the three bodies, which is a frequent occurrence in Saturn's dynamically rich satellite system. 

Not all three-body resonances we encountered so far were of the first order. In some cases when Titan was one of the three bodies involved in the resonance, we observed capture in a second-order three-body eccentricity-type resonance. Since the Hamiltonian of a three-body resonance involves the product of the mass of all three satellites \citep{nes98}, it is logical that resonances involving Titan with its dominant mass would be proportionally stronger, and may stay dynamically relevant even when multiplied by an additional factor of $e$.

\begin{figure*}
\epsscale{.8}
\plotone{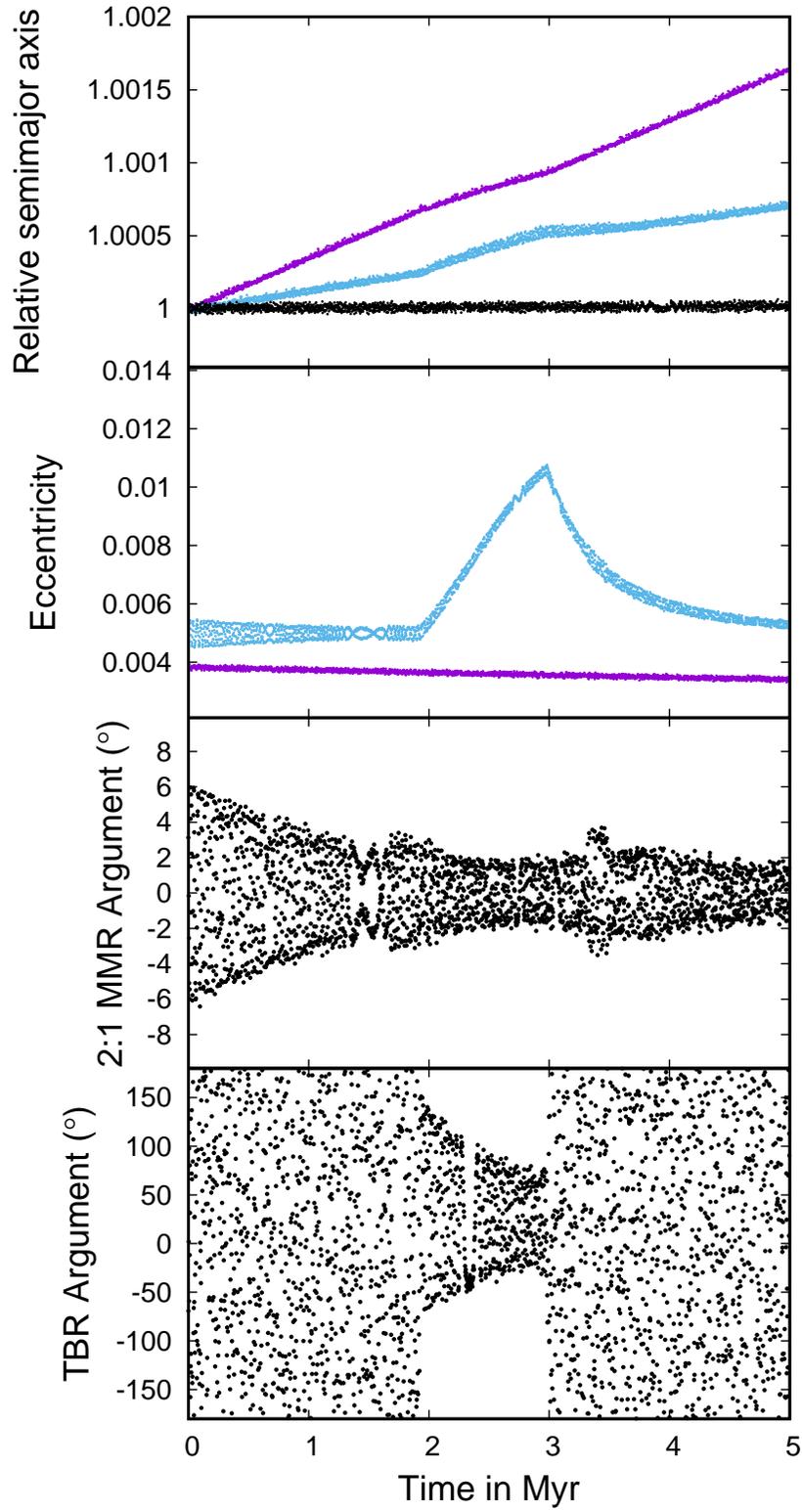}
\caption{A simulations of capture into an isolated three-body resonance between Enceladus (light blue), Tethys (magenta) and Titan (black). The three-body resonance has the argument $7 \lambda_E - 10 \lambda_{\Theta} + 3 \lambda_T + \varpi_E - \varpi_T$ (bottom panel), making it a second-order TBR. Enceladus is also in 2:1 MMR with Dione (second panel from the bottom), which does not participate in the three-body resonance. While the resonance acts to decrease the eccentricity of Enceladus, Enceladus actually becomes more eccentric as it is pushed deeper into the resonance with Dione. This resonance may have happened within the past Myr, opening the possibility that Enceladus's eccentricity and current heating are not currently in equilibrium.\label{enc}}
\end{figure*}

Figure \ref{enc} shows a potentially very recent example of a three-body resonance involving Enceladus, Tethys and Titan. The resonant argument is  $7 \lambda_E - 10 \lambda_{\Theta} + 3 \lambda_T + \varpi_E - \varpi_T$, with the positive sign before $\varpi_E$ indicating that this resonance (considered in isolation) should decrease the eccentricity of Enceladus over time, and would therefore lead to its own breaking. However, Enceladus is concurrently in a 2:1 MMR with Dione, which has a greater effect on the eccentricity of Enecaldus. As the TBR makes Enceladus move deeper into the resonance with Dione, the resulting eccentricity increase due to their 2:1 MMR swamps any direct decrease due to the three-body resonance. The TBR should also increase Titan's eccentricity, but the effect is too small to observe. The resonance breaks within a Myr, apparently due to the influence of Mimas. This resonance should have been encountered within the last 1~Myr for our nominal parameters of Saturn ($Q/k_2 \approx 5000$), opening the possibility that the eccentricity of Enceladus is not in equilibrium, but is still recovering from this resonance. We do note that the probability of capture into this resonance appears to be 20\% at most, making it likely that it was crossed without capture. Despite the presence of the Enceladus-Dione resonance in Fig. \ref{enc}, this TBR can be considered ''isolated'', as neither Tethys nor Titan are in any two-body commensurabilities with any of the other other inner moons, and Enceladus is only in a two-body commensurability with Dione. Therefore, this is a case of a two-body MMR and an isolated TBR happening at the same time, rather than it being a combination or a chain of multiple two-body resonances. 

\section{Discussion and Conclusions}\label{sec:conc}

 Our detection of numerous dynamically significant three-body resonances means that the evolution of the real system cannot be reduced to a handful of two-body MMR passages. In particular, there is a possibility that the eccentricity of Mimas was excited by a three body resonance, possibly the semisecular one shown in Fig. \ref{semisec}, or an  isolated one. Resonances involving Enceladus and Tethys (like those shown in Figs. \ref{chain} and \ref{enc}) could have accelerated the evolution of Enceladus and changed the parameters of its capture into the current resonance with Dione. Additionally, the chronology of the past resonances between Tethys, Dione and Rhea that \citet{cuk16} used to constrain the age of the system now becomes more complicated, as the TBR shown in Fig. \ref{tbr1} slows down the evolution of Tethys but accelerates that of Dione. 


The fact that TBRs are only prominent in numerical simulations using multi-Gyr orbital evolution timescales makes it challenging to study the system's history. Two-body resonances are relatively few in number and their effect can be (to some extent) modelled analytically \citep{mey08, tia20}; in contrast, the large number and complex dynamics of the three-body resonances makes it harder to predict their outcomes. While we were able to automate the search for the locations of first and second- order three-body resonances among the mid-sized moons of Saturn, we were not able to predict the relative strength of these resonances with any confidence without running direct numerical integrations. In other contexts three-body resonances have been studied analytically with good success \citep{qui11, cha18, pet20, pet21}, but more work is needed to determine if a practicable model of TBRs among Saturnian moons can be constructed.

One aspect of significant three-body resonances among Saturnian moons that has stayed consistent in all our simulations is that they do not affect orbital inclinations. D'Alembert's rules dictate that inclination-type resonances always have to be at least of the second order \citep{md99}. Therefore, any inclination-type TBRs would necessarily already be weaker than first-order eccentricity-type ones. However, the order of the resonance by itself is not the only cause of the lack of inclination-type three-body resonances, as we do find capture into {\it second-order} eccentricity-type TBRs in cases when one of the three moons involved in Titan (Fig. \ref{enc}). Our favored interpretation is that strong three-body resonances are combination of first- and zeroth-order two-body terms in the disturbing function, as established by \citet{qui11}. Second order eccentricity-type resonances can result from interaction between two first-order two-body resonant terms. However, inclination-type TBRs would need interactions between a second order and a zeroeth order two-body terms. It appears that, at least in the mass regime of Saturn's satellites, second-order two-body terms are relatively weak and cannot be combined into a three-body resonance capable of capture.

This Letter is only the first report on the existence of three-body resonances capable of resonant capture in the Saturnian system, and it is too early to fully evaluate the importance of these dynamical features for the system's history. However, several direct implications are already clear. First, there is a larger number of opportunities for the moons' eccentricity excitation, and therefore tidal heating, than expected just from the distribution of two-body MMRs. Second, there is the associated complication to modeling the moons' past orbital evolution, as three-body resonances can lead to situations in which an exterior moon (e.g. Tethys) is accelerating the orbital evolution of an interior moon (e.g. Enceladus). Finally, the importance of three-body resonances has conclusively proven that models including only a limited number of two-body resonances cannot capture the full dynamical history of Saturnian moons. Either direct numerical simulations or analytical models including TBRs will be necessary in order to reconstruct this system's complex history.

\acknowledgments

We thank an anonymous reviewer for suggestions that greatly improved the paper. M\'C and MEM are supported by NASA Solar System Workings Program award 80NSSC19K0544.

\bibliography{refs_TBR}{}
\bibliographystyle{aasjournal}

\end{document}